\documentclass[a4paper,11pt]{article}
\pdfoutput=1
\usepackage[a4paper, total={17cm, 25.1cm}]{geometry}
\usepackage{fancyhdr}
\usepackage{epsfig,graphics}
\fancypagestyle{plain}{
}
\begin{document}
%%%%%%%%%%%%%%%%%%%%%%%%%%%%%%%%%%%%%%%%%%%%%%%%%%%%%%%%%%%%%%%%%%%%%%%%%%%%
%%%% Real
\def\R{\mbox{I\hspace{-.15em}R}}

%%%%%%%%%%%%%%%%%%%%%%%%%%%%%%%%%%%%%%%%%%%%%%%%%%%%%%%%%%%%%%%%%%%%%%%%%%%%
%%%% Title

\title{\bf Refinements in the Jungle Universes}

\author{
Alicia \textsc{Simon-Petit} - Han-Hoe \textsc{Yap} - J{\'e}r{\^o}me \textsc{Perez}
\\
\small{Applied Mathematics Departement of ENSTA ParisTech, University of Paris-Saclay}
\\
\small{ alicia.petit@ensta-paristech.fr, 
     han-hoe.yap@ensta-paristech.fr, 
     jerome.perez@ensta-paristech.fr
}
}

\date{January 15, 2016}

\thispagestyle{empty}
\maketitle

\begin{abstract}
How can effective barotropic matter emerge from the interaction of cosmological fluids 
in an isotropic and homogeneous cosmological model ? 

The dynamics of homogeneous and isotropic Friedmann-Lema{\^i}tre-Robertson-Walker 
universes is a natural special case of generalized Lotka-Volterra systems where each 
of the universe’s fluid components can be seen as a competitive species 
in a predator-prey model. (Jungle universe : \cite{2014GReGr..46.1753P})

In addition to numerical simulations illustrating this behaviour among the barotropic 
fluids filling the universe, we analytically pinpoint that effective time-dependent 
barotropic indices can arise from a physical coupling between those fluids whose 
dynamics could then look like that of another type of cosmic fluid, 
such as a cosmological constant.

Since the nature of dark energy is still unknown, this dynamical 
approach could help understanding some of the properties of dark matter 
and dark energy at large cosmological scales.

%\textbf{Keywords.} Cosmology, dynamical systems, Jungle Universe, cosmological coupling,
%interaction, dark energy
\end{abstract}
%%%%%%%%%%%%%%%%%%%%%%%%%%%%%%%%%%%%%%%%%%%%%%%%%%%%%%%%%%%%%%%%%%%%%%%%%%%%

%%%% Paper body

\section*{Introduction}

\indent Einstein's general relativity for gravitation has led him to study the dynamics
of the universe. His cosmology describes an isotropic, homogeneous and static universe, 
while the current $\Lambda$CDM model includes a possible accelerated 
expansion supported by the observation of distant
supernovae~\cite{1998AJ....116.1009R,1997AAS...191.8504P} and the cosmic
microwave background~\cite{2015arXiv150201589P}. 
This late-time cosmic acceleration could be explained by dark energy whose nature 
still remains undetermined.

Several possible explanations have been proposed to account for this acceleration,
from modifications of gravity - with $f(R)$ gravity, scalar-tensor theories, braneworlds - 
to new cosmological fluids such as generalised Chaplygin gas, scalar field with
various couplings with matter, or more naturally non-gravitational couplings between
the constituents of the universe~\cite{amendola2010dark}. But the form of these 
time-dependent~\cite{2014PhRvL.113r1301S} and often linear~\cite{2012PhLB..714....6A} 
interactions generally 
lacks physical justifications, see~\cite{2007PhRvD..75d3515C,2006PhRvD..73j3520B} 
for exact solutions with interacting fluids and 
physical discussion on energy exchanges.

After a reformulation of the standard cosmological model in terms of the density 
proportion of the constituents filling the universe,  
we study a natural quadratic coupling between
cosmological fluids suggested by the intrinsic Lotka-Volterra structure of the fields
equations. This coupling leads to an effective dynamical 
behaviour~\cite{2014GReGr..46.1753P}. 
We eventually propose the basis
for a new interpretation of the acceleration of the universe. 

\section{The standard universe is a generalised predator-prey system}

The dynamics of homogeneous and isotropic Friedmann-Lema{\^i}tre-Robertson-Walker universes 
is a special case of generalised Lotka-Volterra system where the competitive species are the 
barotropic cosmological fluids filling the universe, as it will be underlined in this section. 
\\

The field equations of the standard cosmological model can be written as
\begin{equation}
     \label{eq:FirstFLeqSemiGeneral}
         \;\;\;
           \frac{\ddot a}{ a}
           \,\,\,\,
           = \,
           - \, \frac{4\pi G_N}{3}
                  \sum_{i=1}^{n-1}
                          \left( \rho_i+\frac{3p_i}{c^2} \right)    
          +  \frac{c^2}{3} \, \Lambda
     %\tag{\FirstFLeqSemiGeneral}
\end{equation}
\begin{equation}
     \label{eq:SndFLeqSemiGeneral}
        \phantom{.}
         H^2 
         \,\,
           =
         \,\, 
          \frac{8 \pi G_N}{3} \, \sum_{i=1}^{n-1} \rho_i
          + \frac{c^2}{3} \, \Lambda
          -  \frac{c^2}{a^2} k .
          \,\,\:\:\,
     %\tag{\SndFLeqSemiGeneral}
\end{equation}
Let us consider barotropic fluids with equation of state
$$
   \mathrm{for } \,\, i = 1, \ldots, n+1,
   \,\,\,\;\;\;\;\;\;\;\;\;
   p_i = \omega_i \rho_i \, c^2
   \,\,\,\;\;\;\;\;\; 
   \,\,\,\;\;\;\;\;\;
   \,\,\,\;\;\;\;\;\;
   \,\,\,\;\;\;\;\;\;
   \,\,\;\;\;\;\;\;\;\:\,\,
$$
where each fluid $i = 1, \ldots, n$ can be baryonic matter, radiation, dark matter, dark energy, 
etc.
In the previous equations, $i=n$ would be associated to the cosmological constant
$\Lambda$ with 
$
   \rho_\Lambda = \frac{\Lambda c^2}{8 \pi G_N}
$
and
$
   \omega_\Lambda = -1,
$
while 
$i=n+1$ could be associated to curvature introducing 
$
   \rho_k
    =
   - \frac{3 c^2 k}{8\pi G_N} a^{-2}
$
and
$
   \omega_k = - \frac{1}{3}.
$

A change of variable for the time parameter, $\lambda=\ln(a),$ 
and a reformulation in terms of relative abundances $\Omega_i,$ instead of densities 
$\rho_i,$ lead to the following dynamics for each independent
cosmological fluid governed by a continuity equation of type 
$
   \dot {\rho_i} = - 3 H (\rho_i+\frac{p_i}{c^2})
$
:
$$
   \,
   \mathrm{for } \,\, i = 1, \ldots, n+1,
   \, \; \; \;
   \frac{\mathrm{d}\Omega_i}{\mathrm{d}\lambda}
   =
   \Omega_i \left[
                        - (1+3\,\omega_i) + \sum_{j=1}^n (1+3\,\omega_j) \, \Omega_j
                    \right]
$$
which reads as a predator-prey system, with a community matrix $\textbf{A}$ 
and a capacity vector $\textbf{r}$
$$
 \begin{array}{c}
   \frac{\mathrm{d}}{\mathrm{d}\lambda} 
       \left[
          \begin{array}{c}
             \Omega_1 \\
             \vdots    \\
             \Omega_n \\
             \Omega_k 
          \end{array}
       \right]
   =
   \mathrm{diag}
   \left[ 
      \begin{array}{c}
         \Omega_1       \\
         \vdots         \\
         \Omega_n     \\
         \Omega_k     
      \end{array}
   \right]
   \left(
      \left[
         \begin{array}{cccc}
            1+3 \, \omega_1 & \ldots & 1+3 \, \omega_n & 0     \\
            \vdots               & \ddots & \vdots               & \vdots \\
            1+3 \, \omega_1 & \ldots & 1+3 \, \omega_n & 0     \\
            1+3 \, \omega_1 & \ldots & 1+3 \, \omega_n & 0     
         \end{array}
      \right]
      \left[
         \begin{array}{c}
            \Omega_1 \\
            \vdots    \\
            \Omega_n \\
            \Omega_k
         \end{array}
      \right]
      +
      \left[
         \begin{array}{c}
            - \, (1+3 \, \omega_1) \\
            \vdots    \\
            - \, (1+3 \, \omega_n) \\
            - \, (1+3 \, \omega_k) 
         \end{array}
      \right]
   \right)
\\
\phantom{\small{..........................................}}
\underbrace{
  \phantom{ \small{.....................}
         \left[
            \begin{array}{ccccc}
               0  & 3 (\omega_2 - \omega_1) 
            \end{array}
         \right]
  }
}_{\mathrm{\textbf{A} \,\, matrix}}
\phantom{\small{....................}}
\underbrace{
  \phantom{ \small{..}
         \left[
            \begin{array}{c}
              3 (\omega_2 - \omega_1) 
            \end{array}
         \right]
  }
}_{\mathrm{\textbf{r} \, \, vector}}
\end{array}
$$
in
$\R^{n+1}.$
The search for equilibria in a Lotka-Volterra system governing the evolution of
$n+1$ independent fluids, with a $(n+1)^{\mathrm{th}}$ fluid made of curvature with index $k,$ 
consists in solving
$
   \dot \Omega
   =
   0
   =
   \mathrm{diag}(\Omega) \, (\textbf{A}\,\Omega+\textbf{r}),
$
see e.g.~\cite{9780511524660} chap. 4.
If all fluids interact with each other only gravitationally, then the system
$(\textbf{A}\,\Omega+\textbf{r})=0$ has no solution, since the capacity vector does not 
lie in the image of the community matrix.
The flat Minkowski spacetime
is the solution of $\mathrm{diag}(\Omega)=0,$ while all other equilibria
correspond to universes containing a single fluid $j$, i.e. when $\Omega_j\neq 0$ 
but $\Omega_{i\neq j} = 0,$ then $\mathrm{diag}(\Omega)$
is a zero divisor matrix.

\begin{figure}[h]
   \centering
   \includegraphics[width=0.36\linewidth]{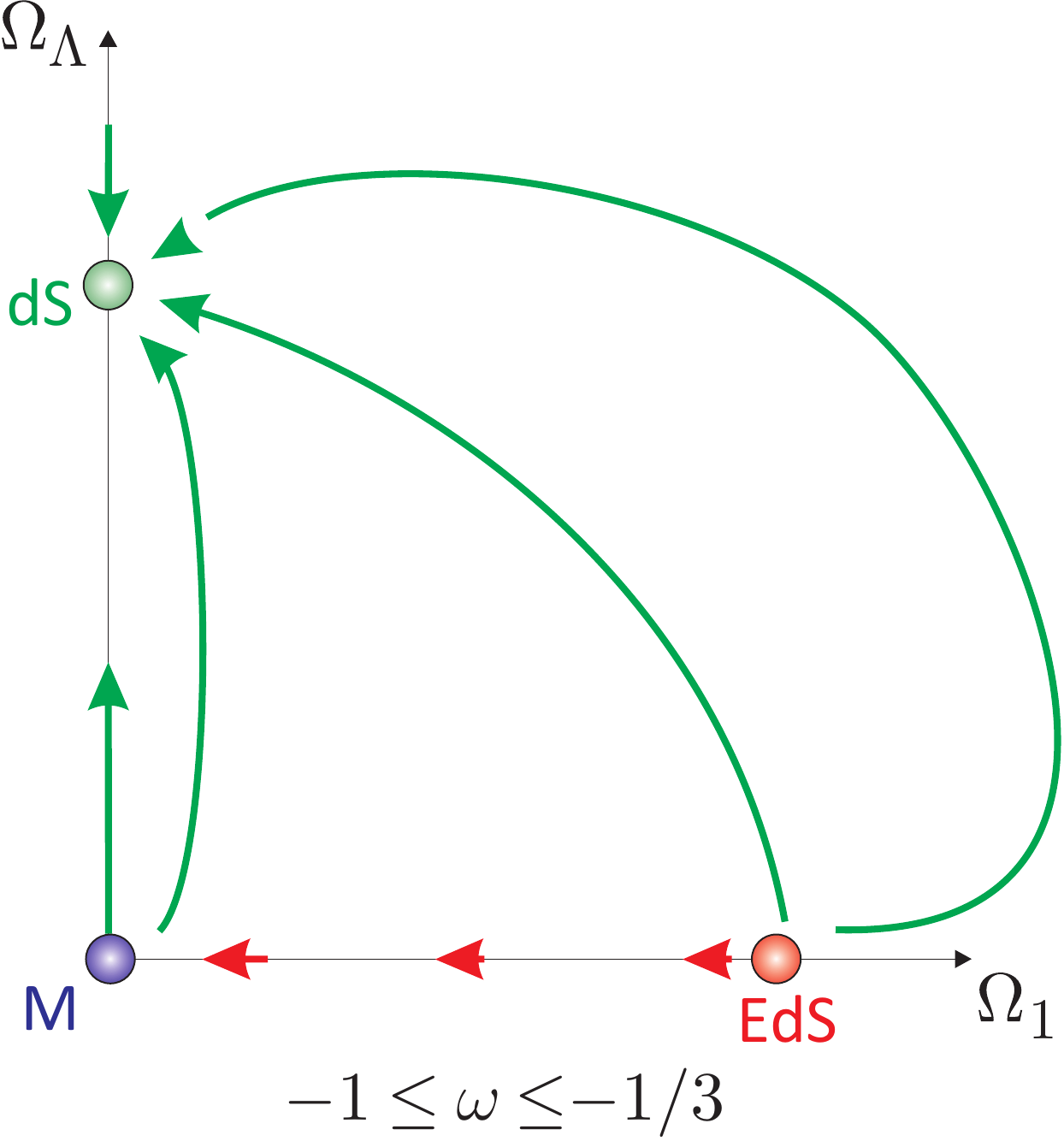}
   $\,\;\,\,\,\,\,\;\;\;\;\;\,$
   \includegraphics[width=0.36\linewidth]{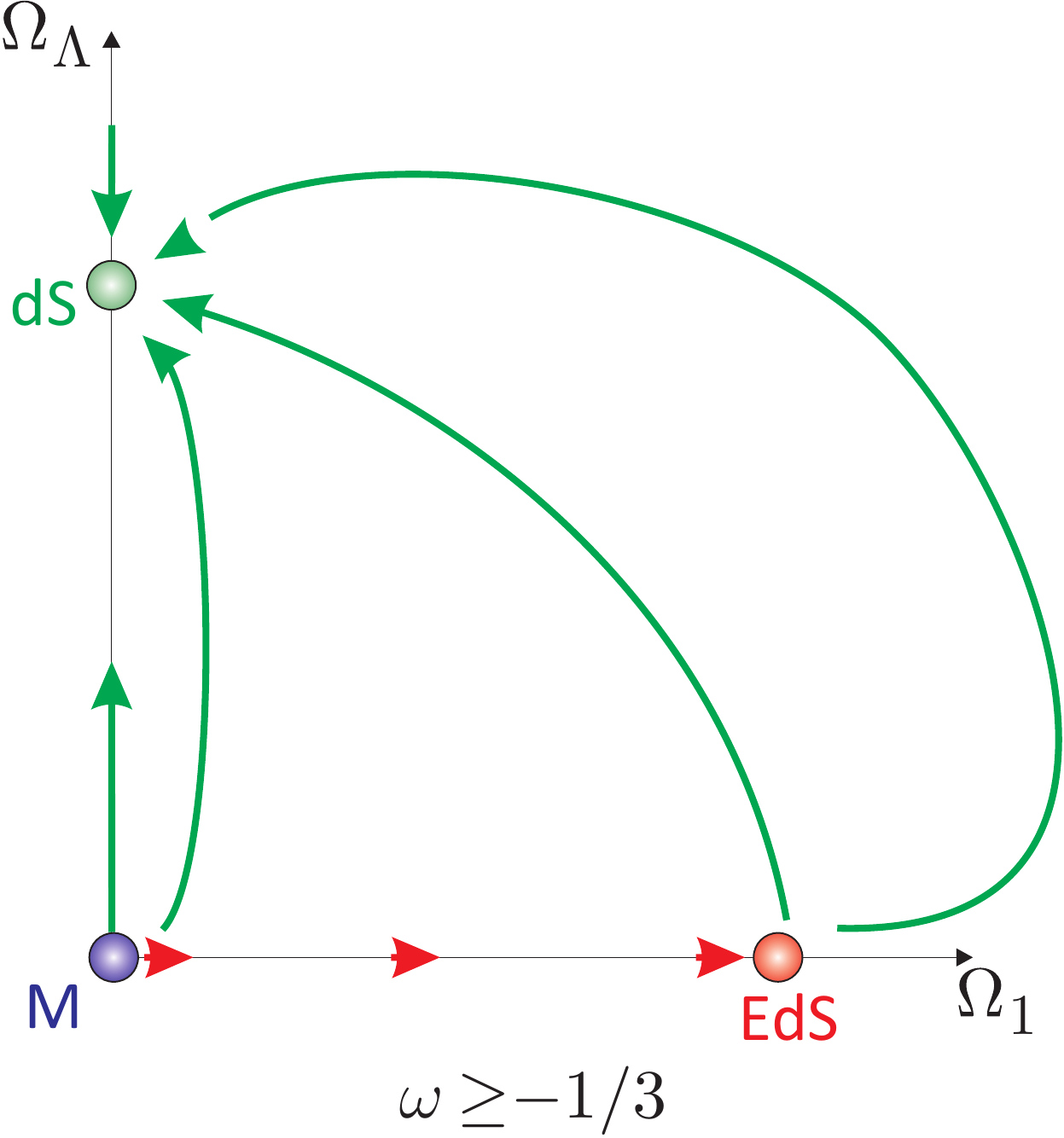}
   \caption[Dynamics in the $(\Omega_m,\Omega_\Lambda)$-plane.]{ 
                 Dynamics in the $(\Omega_1,\Omega_2)=
                                             (\Omega_1,\Omega_\Lambda)$-plane, 
                 considering a species 1, a cosmological constant and curvature. 
                 Time increases when following the arrows and we define 
                 $\omega=\omega_1$ 
                 (e.g. $\omega=\omega_m =0$ 
                 for baryonic matter in the $\Lambda$CDM model).
                 Common universes such as Einstein-de Sitter (EdS), de Sitter (dS) or Milne (M)
                 universes are equilibria whose stability depends on the nature of fluid $1,$
                 i.e. on the value of $\omega.$  
                 }
   \label{fig:Standard2DG} 
\end{figure}

The usual asymptotic states of FLRW universes, 
such as Einstein-de Sitter ($\Omega_m\neq 0$), 
de Sitter ($\Omega_\Lambda \neq 0$) or
Milne  ($\Omega_k \neq 0$) universes, thus appear as particular equilibria between cosmic species, where generally
one species dominates the others, see Fig.~\ref{fig:Standard2DG}.
These equilibria correspond to $\Omega$ vectors with only one non-zero value
which equals 1 because of (\ref{eq:SndFLeqSemiGeneral}). 
Consequently, in the absence of non-gravitational interactions, 
they must lie on the axes of $(\Omega_i,\Omega_j)$-planes 
for all $i,j=0, \ldots, n+1.$

\section{Jungle Universes : cooperation and competition among cosmic fluids}

A universe without any interaction between cosmological fluids
apart from a gravitational one seems a little awkward, whereas a natural coupling
between cosmic fluids leads to a much richer dynamics. 

\begin{figure}[h]
   \centering
   \includegraphics[width=0.9\linewidth]{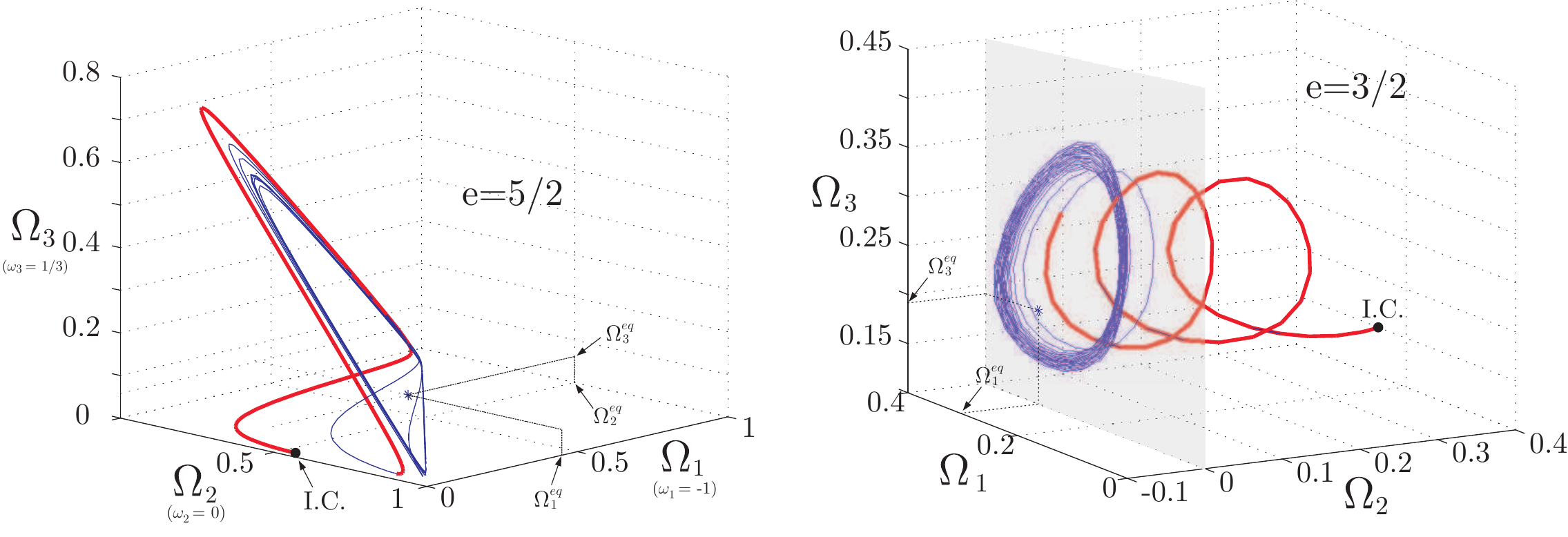}
   \caption[Evolution of the three coupled fluids]{
                  Evolution of the three coupled relative abundances, in the 3D phase space, 
                  with coupling constants : 
                  $\epsilon_{12}=\epsilon_{13}=\epsilon_{23}=\mathrm{e}.$
                  The beginning of the orbits are marked by a bold red line. 
                  Initial conditions are indicated by I.C. and a black dot.
                  Relevant equilibria are indicated by stars.
                 }
   \label{fig:JungleTriads}  
\end{figure}

Various types of coupling, most of the time linear in the densities~\cite{2012PhLB..714....6A}, 
have been proposed by several authors, see e.g.~\cite{2014PhRvL.113r1301S},
but very few with non-linear interactions have been studied~\cite{PhysRevD.83.023528, 2003MPLA...18..831M, 2010EPJC...69..509M}. Except for scalar fields, the non-linearity is introduced in the equations of state~\cite{2006PhRvD..74b3523A} but rarely 
as an interaction between fluids~\cite{2012AIPC.1471...51Z,2015MPLA...3050134M}. 
A 
natural non-linear coupling preserving the intrinsic Lotka-Volterra structure previously
mentioned has been studied in~\cite{2014GReGr..46.1753P}. 
This ``jungle coupling",   
defined by
$
  Q(\epsilon_{ij}) 
  = 
  \epsilon_{ij} \, \Omega_i \, \Omega_j,
$ 
naturally vanishes in the absence of one of the interacting fluids,
and the interaction rate increases with the coupled densities.
Interactions which are proportionnal to the product of these densities 
also provide one of the best observational fits for holographic dark energy models 
coupled to dark matter~\cite{2010EPJC...69..509M} and are the bests, with 
respect to linear couplings, to alleviate the coincidence problem with common
time-dependent equations of state~\cite{2008JCAP...06..010H}.
The quadratic coupling term
$
  Q(\epsilon_{ij}) 
$
added to and substracted from the conservation equations of the interacting fluids
$i$ and $j$
$$
             \frac{\mathrm{d}\Omega_{i/j}}{\mathrm{d}\lambda}
             = \,
             \Omega_{i/j} \left[
                                              - (1+3\,\omega_{i/j}) 
                                              + \sum_{l=1}^n (1+3\,\omega_l) \, \Omega_l     
                                       \right]
             \begin{array}{c}+/-\end{array} \, Q(\epsilon_{ij})
$$
partially breaks the degeneracy of the community matrix $\textbf{A}.$
Equilibria may be located outside the axes of $(\Omega_i, \Omega_j)$-planes.
The associated dynamics can contain limit cycles (Fig.~\ref{fig:JungleTriads}), 
universes with several fluids
in equilibrium, and even chaos which has been conjectured for more than four
fluids in interaction. 
The dynamical behaviours in this so-called Jungle Universe
could answer some questions such as the coincidence problem.

Further dynamical properties are explained in~\cite{2014GReGr..46.1753P}.
In the last section, we will employ the time evolution of the dynamical behaviour 
of interacting cosmological fluids to look for the possibility of an effective dark energy.
\\

\section{Camouflage in the jungle : could dark energy emerge from the jungle coupling ?}

The observed accelerated expansion of the universe suggests the existence of a cosmological
constant or at least dark energy of unknown nature. Could a coupling among cosmic fluids
lead to a similar behaviour instead ?

Using the formalism of Jungle Universes, we suggest hereafter that the acceleration 
of the expansion could result from
a special interaction between fluids of known properties.
The interaction term in the continuity equation of a fluid $i$ reads
$$
      \dot {\rho_i} 
      = 
      - 3 H (\rho_i+\frac{p_i}{c^2})
      + \sum_{j=1}^{n}  \epsilon_{ij}  H \, \Omega_j \, \rho_i.
$$
It actually modifies 
its equation of state which then describes a barotropic fluid with an 
effective time-dependent barotropic index
$$
       \omega_i^{\mathrm{eff}}
       = 
       \omega_i - \sum_{j=1}^{n}  \frac{\epsilon_{ij}}{3} \,\,  \Omega_j.          
$$
Consequently a dominant negative term can appear in the second member 
of (\ref{eq:FirstFLeqSemiGeneral}),
which is equivalent to an acceleration $\ddot a >0.$

\begin{figure}[h]
   \centering
   \includegraphics[width=0.44\linewidth]{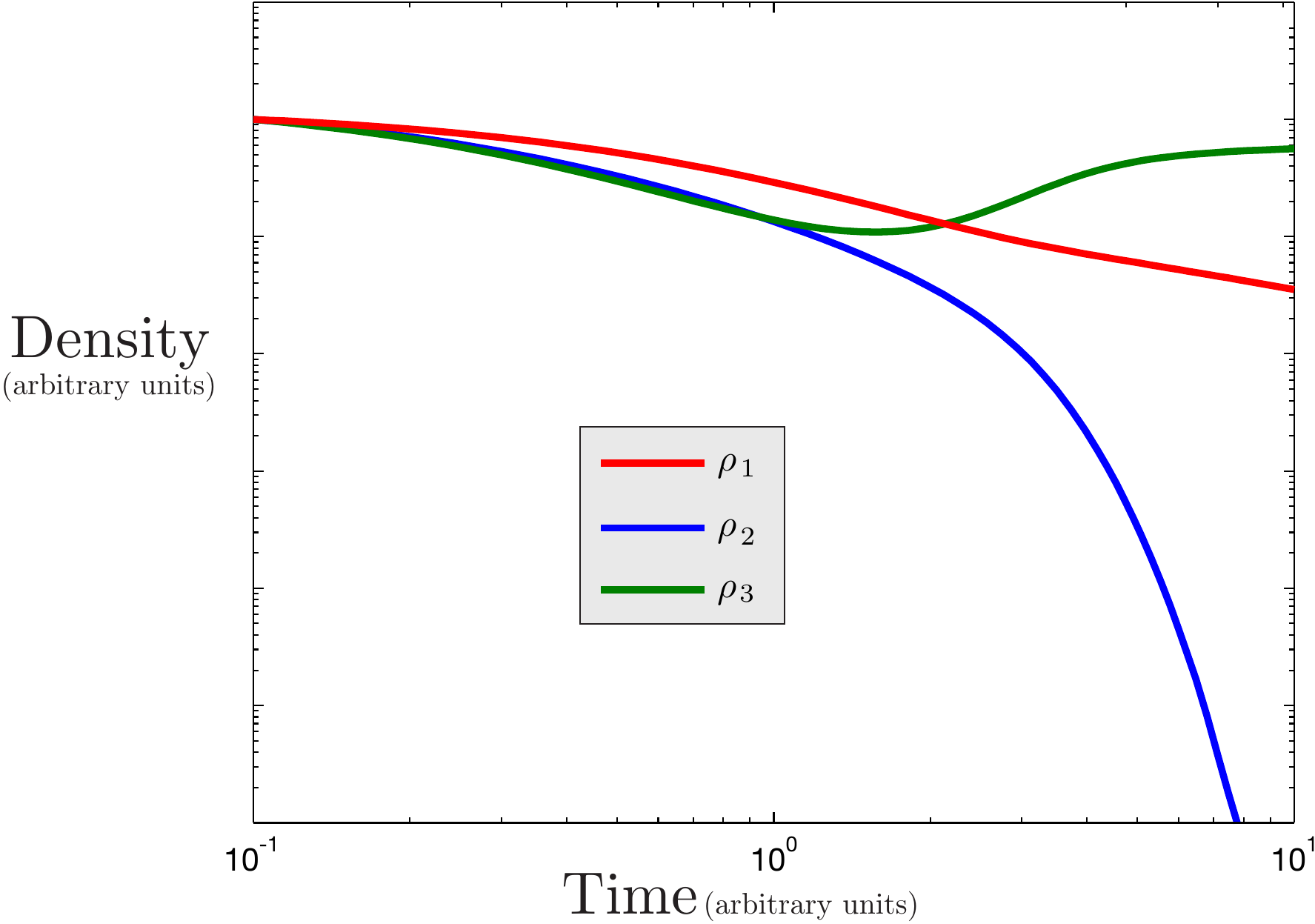}
   \,\,\,\,\,
   \includegraphics[width=0.46\linewidth]{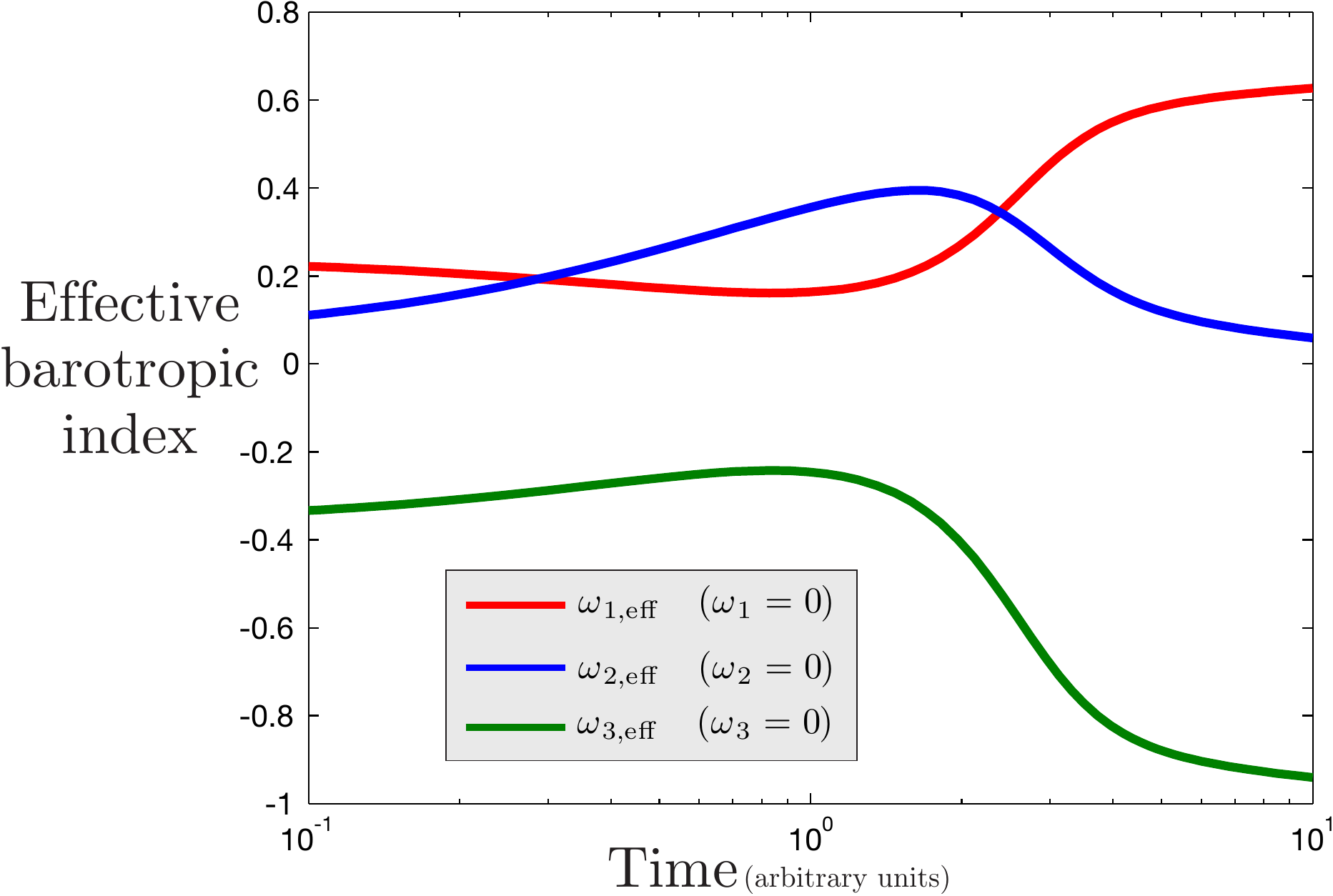}
   \,\,
   \caption[Interaction between three dark matter fluids]{
                 Interaction between three dark matter fluids : evolution of their density
                 and effective barotropic index. Coupling constants : 
                $\epsilon_{12}=-2$ and $\epsilon_{23}=-3.$
                 }
   \label{fig:EffectiveEvol3DM}  
\end{figure}

As an illustration, consider three dark matter fluids with the first two and last two interacting
with each other. 
Indexing them from 1 to 3, with $\epsilon_{12}=-2$ and $\epsilon_{23}=-3,$
the mutual interactions make the third fluid
asymptotically dominate the others in terms of density with an effective barotropic index
close to $-1,$ see Fig.~\ref{fig:EffectiveEvol3DM}. Thus fluid 3 dynamically behaves nearly as a cosmological constant. In the same way, the first fluid looks like a perfect gas while the second
one still behaves as a dark matter. 
\\

Couplings between barotropic cosmological fluids change their observed behaviour 
and can influence the global dynamics of the universe. The acceleration of the 
universe could then be explained in a natural way without introducing  
unknown types of energy.

\section*{Conclusion}
The natural generalised Lotka-Volterra structure of the evolution of the universe
constituent densities has enabled us to describe the isotropic and homogenous
universe as a generalised predator-prey system. The effective barotropic indices
that emerge from the natural quadratic jungle coupling between cosmological fluids 
induce various dynamics of the evolution of the scale factor and fluid densities, 
from limit cycles to possible chaos. 
We have 
made a simulation of an effective cosmological constant 
from a coupling between three dark matter fluids. 
Similar interactions could explain the observed acceleration of the universe. 
Comparisons with cosmological data are therefore planned to be processed.
\\

%\textbf{Keywords.} Cosmology, dynamical systems, Jungle Universe, cosmological coupling,
%interaction, dark energy
%\\

\textbf{Acknowledgment}
This work is supported by the "IDI 2015" project funded by the IDEX Paris-Saclay, 
ANR-11-IDEX-0003-02.

%%%% References

%%%%%%%%%%%%%%%%%%%%%%%%%%%%%%%%%%%%%%%%%%%%%%%%%%%%%%%%%%%%%%%%%%%%%%%%%%%%

\begin{thebibliography}{99}
\bibitem{1998AJ....116.1009R}
Riess et al. (1998) Astrophysical Journal (1998) \textbf{116}, pp. 1009-1038
\bibitem{1997AAS...191.8504P}
Perlmutter et al. (1997) Bulletin of the American Astronomical Society, \textbf{29}
\bibitem{2015arXiv150201589P}
Planck Collaboration, Ade, P.~A.~R. et al. (2015) Submitted to A\&A.
\bibitem{amendola2010dark}
Amendola, L. and Tsujikawa, S. (2010) Cambridge University Press
\bibitem{2014PhRvL.113r1301S}
Salvatelli, V. et al. (2014) Physical Review Letters, \textbf{113}, 18, p. 181301
\bibitem{2012PhLB..714....6A}
Avelino, P.~P. and da Silva, H.~M.~R. (2012) Phys. Rev. B, \textbf{714}, pp. 6-10
\bibitem{2014GReGr..46.1753P}
Perez, J. et al. (2014) General Relativity and Gravitation, \textbf{46}, p. 1753
\bibitem{2007PhRvD..75d3515C}
Clifton, T. and Barrow, J.~D. (2007) Phys. Rev. D., \textbf{75}, 4, p. 043515
\bibitem{2006PhRvD..73j3520B}
Barrow, J.~D. and Clifton, T. (2006) Phys. Rev. D., \textbf{73}, 10, p. 103520
\bibitem{9780511524660}
Wainwright, J. and Ellis, G.~F.~R. (1997) Cambridge University Press
\bibitem{PhysRevD.83.023528}
Lip, Sean Z. W. (2011) Phys. Rev. D, \textbf{83}
\bibitem{2003MPLA...18..831M}
Mangano, G., Miele, G. and Pettorino, V. (2003) Modern Physics Letters A, \textbf{18}, pp. 831-842
\bibitem{2010EPJC...69..509M}
Ma, Y.-Z., Gong, Y. and Chen, X. (2010) European Physical Journal C, \textbf{69}, pp. 509-519
\bibitem{2006PhRvD..74b3523A}
Ananda, K.~N. and Bruni, M. (2006) Phys. Rev. D., \textbf{74}, 2, p. 023523
\bibitem{2012AIPC.1471...51Z}
Zimdahl, W. (2012) American Institute of Physics Conference Series, \textbf{1471}, pp. 51-56
\bibitem{2015MPLA...3050134M}
Mahata, N. and Chakraborty, S. (2015) Modern Physics Letters A, \textbf{30}, p. 1550134
\bibitem{2008JCAP...06..010H}
He, J.-H. and Wang, B. (2008) Journal of Cosmology and Astroparticle Physics, \textbf{6}, p. 10.
\end{thebibliography}
\end{document}